\documentclass[aps,preprint,amssymb,showpacs,tightenlines]{revtex4}

\usepackage{amsmath}
\usepackage{amssymb}
\usepackage{amsthm}
\usepackage{pstricks,pst-node,pst-plot}
\newpsobject{showgrid}{psgrid}{subgriddiv=1, griddots=5, gridlabels=6pt}
\usepackage{graphicx}

\newcommand{\beq}{\begin{equation}}
\newcommand{\eeq}{\end{equation}}
\newcommand{\beqa}{\begin{eqnarray}}
\newcommand{\eeqa}{\end{eqnarray}}
\newcommand{\beg}{\begin{gather}}
\newcommand{\eeg}{\end{gather}}

\begin{document}

\title{Decay of the free-theory vacuum of scalar field theory in de~Sitter spacetime in the interaction picture}

\author{Atsushi Higuchi}

\affiliation{Department of Mathematics, University of York,
Heslington, York YO10 5DD, United Kingdom\\ email:
ah28@york.ac.uk}

\date{23 December, 2008}

\begin{abstract}
A free-theory vacuum state of an interacting field theory, e.g.~quantum gravity, is unstable at tree level in general due to spontaneous emission of Fock-space particles in any spacetime with no global timelike Killing vectors, such as de~Sitter spacetime, in the interaction picture.
As an example, the rate of spontaneous emission of Fock-space particles is calculated in $\varphi^4$ theory in de~Sitter spacetime.  It is possible that this apparent spontaneous emission does not 
correspond to any physical processes because the states are not evolved by the true Hamiltonian in the interaction picture.  Nevertheless, the constant spontaneous emission of Fock-space particles in the interaction picture clearly demonstrates that the in- and out-vacuum states are orthogonal to each other as emphasized by Polyakov and that the
in-out perturbation theory, which presupposes some overlap between these two vacuum states, is inadequate.
Other possible implications of apparent vacuum instability of this kind in the interaction picture are also discussed.
\end{abstract}

\pacs{04.62.+v}

\maketitle

\section{Introduction}

Inflationary cosmological models~\cite{Guth,Linde,Steinhardt,Sato1,Sato2,Kazanas}, which assume exponential expansion of the Universe in its early stage, have become very important in modern cosmology (see also \cite{Starobinsky}).  The spacetime in the expansion stage of these models is approximately de~Sitter spacetime, which is the unique maximally symmetric solution to the Einstein equations with a positive cosmological constant (see, e.g., \cite{HawkingEllis}).  Moreover, the Universe at present may approximately be de~Sitter spacetime since it appears to be undergoing accelerated expansion~\cite{Riess,Perlmutter}. (See \cite{Turner} for a recent review.)  For these reasons quantum field theory in de~Sitter spacetime is attracting much attention recently.  The cosmological constant problem~\cite{Weinberg}, the fact that the cosmological constant is much smaller than naturally expected from the Standard Model of particle physics, is another reason for studying quantum field theory in de~Sitter spacetime. 

Some authors have suggested that the vacuum states in quantum field theories, including quantum gravity, may be unstable in de~Sitter spacetime (see, e.g., \cite{Yokoyama,Polyakov}). In particular, Polyakov~\cite{Polyakov} has pointed out that in this spacetime the free-theory vacuum state is generally unstable against spontaneous emission of Fock-space particles at tree level in the interaction picture in any interacting field theory such as general relativity. (We use the phrase ``Fock-space particles" to mean the quanta created by creation operators in a Fock space.  They should not be confused with particles detected, e.g.~by an Unruh-DeWitt detector~\cite{Unruh,DeWitt}.)  In Minkowski spacetime such processes are forbidden due to energy conservation because the vacuum state has the lowest energy.  However, it is clear that the conservation laws in de~Sitter spacetime do not prevent such processes from occurring. (This conclusion should be true in any spacetime without a global timelike Killing vector.) Several authors have pointed out that interacting low-mass scalar fields have infrared-divergent $n$-point functions~\cite{Yokoyama,Polyakov,Sasaki}.  
The relation between these infrared divergences and the spontaneous emission discussed here is not very clear since the latter occurs for interacting fields of any mass and spin.

In this paper we first calculate the rate per unit volume of spontaneous emission of four Fock-space particles in the interaction picture in the theory in which the conformally-coupled massless scalar field interacts through a $\varphi^4$ term. We also present the expression for the emission rate for a scalar field of arbitrary mass to emphasize that this process is not entirely an infrared effect. Then we discuss possible significance of apparent spontaneous emission of this kind in general.  We use the metric signature $+---$ and natural units $\hbar = c = 1$ throughout this paper unless otherwise stated.

\section{Calculation of the spontaneous emission rate}

One can cover half of de~Sitter spacetime by the coordinates $(u,\mathbf{x})$ with the conformally-flat metric of the form
\beq
ds^2 = (Hu)^{-2}(du^2 - d\mathbf{x}\cdot d\mathbf{x})\,. \label{metric}
\eeq
Here, $\mathbf{x}$ is a three-dimensional vector, and the conformal time $u$ decreases from $\infty$ to $0$ towards the future. The constant $H$ is the Hubble constant, which gives the rate of expansion of the space.
The Lagrangian density of the conformally-coupled massless scalar field $\varphi(u,\mathbf{x})$ interacting through the $\varphi^4$ term is
\beq
{\cal L} = \sqrt{-g}\left[\frac{1}{2}(\nabla_\mu \varphi)(\nabla^\mu \varphi) - \frac{1}{12}R\varphi^2 - \frac{\lambda}{4!}\varphi^4\right]\,,
\eeq
where $g= -(Hu)^{-8}$ is the determinant of the metric tensor $g_{\mu\nu}$ and where the scalar curvature is given by $R=12H^2$.  Treating the $\varphi^4$ term as an interaction term, we find that the free field (or the field in the interaction picture), $\varphi_I$, satisfies
\beq
\left(\Box + 2H^2\right)\varphi_I = 0\,.
\eeq
As is well known, this equation can be written as the massless scalar field equation for 
$(Hu)^{-1}\varphi_I(u,\mathbf{x})$ in (part of) Minkowski spacetime with the metric $-du^2+d\mathbf{x}\cdot d\mathbf{x}$.   Hence, the field $\varphi_I(u,\mathbf{x})$ can be expanded as
\beq
\varphi_I(u,\mathbf{x}) =
\int \frac{d^3\mathbf{k}}{(2\pi)^3}\left[ a(\mathbf{k})\frac{Hu}{\sqrt{2k}}e^{iku+i\mathbf{k}\cdot\mathbf{x}}
+ a^\dagger(\mathbf{k})\frac{Hu}{\sqrt{2k}}e^{-iku-i\mathbf{k}\cdot\mathbf{x}}\right]
\eeq
with $k\equiv \|\mathbf{k}\|$,
where the annihilation and creation operators satisfy the usual commutation relations,
\beq
\left[a(\mathbf{k}_1),a^\dagger(\mathbf{k}_2)\right] = (2\pi)^3\delta^3(\mathbf{k}_1-\mathbf{k}_2)\,,
\eeq
with all other commutators vanishing.  The vacuum state $|0\rangle$ is defined by requiring that $a(\mathbf{k})|0\rangle = 0$ for all $\mathbf{k}$.  This state is the standard vacuum state called the Euclidean (or Bunch-Davies) 
vacuum~\cite{GibbonsHawking,BunchDavies}. (This choice of vacuum is implicit in an earlier work of Tagirov~\cite{Tagirov}.)

We define the transition amplitude $\mathcal{A}(\mathbf{k}_1,\mathbf{k}_2,\mathbf{k}_3,\mathbf{k}_4)$ from the free-theory vacuum state $|0\rangle$ to a 4-Fock-space-particle state, 
$$
|\mathbf{k}_1,\mathbf{k}_2,\mathbf{k}_3,\mathbf{k}_4\rangle = a^\dagger(\mathbf{k}_1)a^\dagger(\mathbf{k}_2)a^\dagger(\mathbf{k}_3)a^\dagger(\mathbf{k}_4)|0\rangle\,,
$$
to lowest order in $\lambda$ as
\beq
\mathcal{A}(\mathbf{k}_1,\mathbf{k}_2,\mathbf{k}_3,\mathbf{k_4})
\equiv \int dud^3\mathbf{x}\,\sqrt{-g}\,
\langle \mathbf{k}_1,\mathbf{k}_2,\mathbf{k}_3,\mathbf{k}_4|\mathcal{H}_I(u,\mathbf{x})|0\rangle\,,
\eeq
where
\beq
\mathcal{H}_I(u,\mathbf{x}) = \frac{\lambda}{4!}\left[\varphi_I(u,\mathbf{x})\right]^4\,.
\eeq
We readily obtain
\beq
\mathcal{A}(\mathbf{k}_1,\mathbf{k}_2,\mathbf{k}_3,\mathbf{k}_4)
= \lambda \int_0^\infty du\, \frac{e^{-i(k_1+k_2+k_3+k_4)u}}{4\sqrt{k_1k_2k_3k_4}}(2\pi)^3\delta^3(\mathbf{k}_1+\mathbf{k}_2+\mathbf{k}_3+\mathbf{k}_4)\,.
\label{amp}
\end{equation}
It is possible to integrate over $u$ after introducing a cutoff by letting $k_1+\cdots+k_4 \to k_1+\cdots+k_4-i\epsilon$, where $\epsilon$ is an arbitrarily small positive number.  Nevertheless, we leave this integral as it is until we square the amplitude in order to factor out the infinite spacetime volume from the transition probability to find the transition rate per unit volume. 

First we make the change of variable $u = H^{-1}e^{-Ht}$, where $t$ is the proper time for a geodesic observer with $\mathbf{x}$ constant.  Then the metric (\ref{metric}) takes the form
\beq
ds^2 = -dt^2 + e^{2Ht}d\mathbf{x}\cdot d\mathbf{x}\,. \label{metric2}
\eeq
This form of the metric shows clearly that the space expands with Hubble constant $H$.
The transition probability is
\begin{eqnarray}
P & = & \frac{1}{4!}\int \frac{d^3\mathbf{k}_1}{(2\pi)^3}\frac{d^3\mathbf{k}_2}{(2\pi)^3}
\frac{d^3\mathbf{k}_3}{(2\pi)^3}\frac{d^3\mathbf{k}_4}{(2\pi)^3}
|\mathcal{A}(\mathbf{k}_1,\mathbf{k}_2,\mathbf{k}_3,\mathbf{k}_4)|^2\nonumber \\
& = & 
\frac{\lambda^2}{4!}\int \frac{d^3\mathbf{k}_1d^3\mathbf{k}_2 d^3\mathbf{k}_3d^3\mathbf{k}_4}{(2\pi)^{12}}
\int_{-\infty}^\infty dt_1 \int_{-\infty}^\infty dt_2 \frac{e^{-H(t_1+t_2)}}{16k_1k_2k_3k_4}\nonumber \\
&& \times\exp\left[\frac{i(k_1+k_2+k_3+k_4)}{H}\left(e^{-Ht_1} - e^{-Ht_2}\right)\right]
(2\pi)^3 \delta^3(\mathbf{k}_1+\mathbf{k}_2+\mathbf{k}_3+\mathbf{k}_4) V_c\,,
\end{eqnarray}
where we have interpreted $(2\pi)^3\delta^3(\mathbf{0})$ in the momentum space as the infinite {\em coordinate} volume 
$V_c=\int d^3\mathbf{x}$~\cite{Bjorken}.  Changing the variables again as $T=(t_1+t_2)/2$ and $\tau = t_2-t_1$, we find 
\begin{eqnarray}
P & = & \frac{\lambda^2}{4!}\int \frac{d^3\mathbf{k}_1d^3\mathbf{k}_2d^3\mathbf{k}_3d^3\mathbf{k}_4}{16(2\pi)^9 k_1k_2k_3k_4}
\delta^3(\mathbf{k}_1+\mathbf{k}_2+\mathbf{k}_3+\mathbf{k}_4)\nonumber \\
&& \times \int_{-\infty}^\infty dT \int_{-\infty}^\infty d\tau
\,e^{-2HT}\exp\left[\frac{2i(k_1+k_2+k_3+k_4)}{H}e^{-HT}\sinh \frac{H\tau}{2}\right]V_c\,.
\end{eqnarray}
It is useful to multiply this expression by $\delta(k_1+k_2+k_3+k_4 - K)$ and integrate over $K$ from $0$ to $\infty$ and then change the integration variables from $\mathbf{k}_i$ to $\mathbf{y}_i = K^{-1}\mathbf{k}_i$, $i=1,2,3,4$.  Then, using the result
\beq
\int\frac{d^3\mathbf{y}_1d^3\mathbf{y}_2d^3\mathbf{y}_3d^3\mathbf{y}_4}{y_1y_2y_3y_4}\delta(y_1+y_2+y_3+y_4-1)\delta^3(\mathbf{y}_1+\mathbf{y}_2+\mathbf{y}_3+\mathbf{y}_4) = \frac{\pi^3}{4}\,,
\eeq
where $y_i \equiv \|\mathbf{y}_i\|$, $i=1,2,3,4$, we find
\beq
P  = \frac{\lambda^2}{3(8\pi)^6}\int_{-\infty}^\infty dT \int_0^\infty dK\,K^4\int_{-\infty}^\infty d\tau\,
e^{-2HT}\exp\left(\frac{2iK}{H}e^{-HT}\sinh \frac{H\tau}{2}\right)V_c\,.
\eeq

Now, the metric given by (\ref{metric2}) shows that at time $T$ the proper distance $\ell_p$ between two points $\mathbf{x}_1$ and $\mathbf{x}_2$ is related to the coordinate distance $\ell_c=\|\mathbf{x}_1-\mathbf{x}_2\|$ 
by $\ell_p = e^{HT}\ell_c$.  Thus, the physical wave number vector of the mode with label $\mathbf{k}$ at time $T$ is $e^{-HT}\mathbf{k}$.  This fact motivates the change of variable from $K =k_1+k_2+k_3+k_4$ to
$\kappa = (2K/H)e^{-HT}$, which is roughly the typical wave number of the emitted Fock-space particles in units of the Hubble constant $H$. (A similar change of variables is used in the standard calculation for the response rate of a uniformly accelerated detector in Minkowski spacetime~\cite{BirrellDavies}.) Then the probability $P$ can be written as
\beq
P = \int_{-\infty}^\infty dT\,\Gamma\,V_c e^{3HT}\,,
\eeq
where $V_c e^{3HT}$ can be interpreted as the total {\em proper} volume of the Universe at time $T$, and where $\Gamma$ is interpreted as the emission rate per unit volume and given by
\beq
\Gamma =\frac{\lambda^2H^4}{48(8\pi)^6}\int_0^\infty d\kappa\,\kappa^4
\int_{-\infty}^\infty d\eta\,
\exp\left(i\kappa \sinh \eta\right)\,.
\eeq
Here we have made a further change of variable $H\tau/2 = \eta$.
This integral can be evaluated using standard integrals~\cite{GR} as
\begin{eqnarray}
\Gamma & = & \frac{\lambda^2H^4}{24(8\pi)^6c^3}\int_0^\infty d\kappa\,\kappa^4\,K_0(\kappa) \\
& = & \frac{3\lambda^2H^4}{4(16\pi)^5c^3}\,,
\end{eqnarray}
where we have restored the speed of light $c$ by dimensional analysis. 
(The rate $\Gamma$ is independent of $\hbar$.) Since
$K_0(\kappa) \approx \sqrt{\pi}(2\kappa)^{-1/2}e^{-\kappa}$ for $\kappa \gg 1$,
we find that emission of modes with wavelengths much shorter than the horizon scale, $c/H$, is suppressed and that the emission is dominated by modes with wavelengths comparable to $c/H$. 

As we stated before, the rate is nonzero for a scalar field of arbitrary mass.  Let the field be conformally coupled and have mass $m$.  Define $\nu \equiv \sqrt{1/4 - (mc^2/\hbar H)^2}$.  Then the emission rate per unit volume will be
\begin{eqnarray}
\Gamma_m & = & \frac{\lambda^2 H^4e^{-4{\rm Im}\,\nu}}{3(16\pi)^5c^3}
\int d^3\mathbf{y}_1d^3\mathbf{y}_2d^3\mathbf{y}_3 d^3\mathbf{y}_4\delta^3(\mathbf{y}_1
+ \mathbf{y}_2+\mathbf{y}_3+\mathbf{y}_4)\nonumber \\
&& \times \delta(y_1+y_2+y_3+y_4-1)
\lim_{\epsilon\to 0+}\left|\int_0^\infty dx\,x^{7/2}
e^{-\epsilon x}\prod_{i=1}^4H_\nu^{(1)}(y_ix)\right|^2\,,
\end{eqnarray}
where ${\rm Im}\,\nu = \sqrt{(mc^2/\hbar H)^2-1/4}$ if $mc^2/\hbar H > 1/2$.

\section{Summary and Discussions}

One might be tempted to conclude from the calculation in the previous section that the vacuum state would be unstable in the $\varphi^4$ theory in de Sitter spacetime.  However, since the states are not evolved by the true Hamiltonian in the interaction picture used in our calculation, it is not clear whether or not the apparent spontaneous emission process in this picture provides a good description of a physical process.  Nevertheless, our calculation clearly demonstrates that the in-vacuum state, i.e. the no-Fock-space-particle state in the infinite past, evolves to a state with infinitely many Fock-space particles relative to the out-vacuum state in the infinite future.  Thus, the in- and out-vacuum states are orthogonal to each other as emphasized by Polyakov~\cite{Polyakov}. This means that the in-out perturbation theory is inadequate for the $\varphi^4$ theory and other interacting field theories in de Sitter spacetime since it presupposes some overlap between these vacuum states~\cite{Peskin}.  This point can be illustrated more clearly by a free scalar field theory with a small mass term treated as a perturbation~\cite{unpublished}.

Assuming that the spontaneous emission process in the interaction picture studied in this paper gives a good description of a true physical process, let us consider how it would appear to an inertial observer.  Such an observer would describe her quantum field theory using the symmetry generated by the timelike Killing vector inside her cosmological horizon as the time translation symmetry.  In this description of the field theory the energy can be defined as the conserved quantity corresponding to this Killing vector, and the vacuum state, distinct from the Euclidean vacuum, can be defined as the lowest energy eigenstate. Therefore there cannot be spontaneous emission in this description. 
Now, the Euclidean vacuum is seen as a thermal bath of Gibbons-Hawking temperature $H/2\pi$ in this description of the field theory inside the cosmological horizon~\cite{GibbonsHawking}, and the Fock-space particles in this thermal bath interact with one another. For example, there are scattering processes.  The natural conclusion, therefore, is that spontaneous emission of Fock-space particles in the global description of the field theory is seen by an inertial observer as a process involving some initial Fock-space particles, e.g.~scattering of two Fock-space particles in the thermal bath.  This conclusion is analogous to the well-known fact that when an accelerated detector in Minkowski spacetime absorbs a quantum, it {\em emits} a usual Minkowski particle~\cite{UnruhWald}. (See, e.g., \cite{CHM} for some more examples illustrating how the inertial and accelerated observers describe the same phenomenon differently in Minkowski spacetime.) Given that an inertial observer does not see any spontaneous emission in de~Sitter spacetime, it will be interesting to determine whether or not she sees anything unusual in the Gibbons-Hawking thermal bath. 

It will also be interesting to determine the state to which the free-theory vacuum evolves in these processes in the interaction picture.  Since the Euclidean vacuum state is de~Sitter invariant, it cannot make a transition to a de~Sitter non-invariant state in perturbation theory.  Since there are no de~Sitter invariant states other than the vacuum state in the Fock space of the free theory, it might appear that there would be a contradiction.  However, it is known that one can construct de~Sitter invariant states with infinite norm which nevertheless form a well-defined Hilbert space~\cite{Higuchi1}.   These states were constructed in connection with ``quantum linearization instabilities", which lead to the requirement that all physical states be de~Sitter invariant~\cite{Moncrief1,Moncrief2} (see also \cite{Higuchi2}).  (For recent work on quantum linearization instabilities of de~Sitter spacetime see \cite{Losic,Marolfetal,Marolfetal2}.)  
It is natural to speculate that the free-theory vacuum state makes a transition to one of these de~Sitter invariant states in an interacting field theory.  

\acknowledgments

We thank Don Marolf for helpful comments on an earlier version of this paper, Yen Cheong Lee for useful discussions and Professor Starobinsky for useful correspondence.



\begin{thebibliography}{99}

\bibitem{Guth} A.~H.~Guth, Phys.~Rev.~D~{\bf 23}, 347 (1981).

\bibitem{Linde} A.~D.~Linde, Phys.~Lett.~{\bf B108}, 389 (1982).

\bibitem{Steinhardt} A.~Albrecht and P.~J.~Steinhardt, Phys.~Rev.~Lett.~{\bf 48}, 1220 (1982).

\bibitem{Sato1} K.~Sato, Mon.~Not.~Roy.~Astron.~Soc.~{\bf 195}, 467 (1981).

\bibitem{Sato2} K.~Sato, Phys.~Lett.~{\bf 99B}, 66 (1981).

\bibitem{Kazanas} D.~Kazanas, Astrophys.~J.~{\bf 241}, 159 (1980).

\bibitem{Starobinsky} A.~A.~Starobinsky, Phys.~Lett.~{\bf B91}, 99 (1980).

\bibitem{HawkingEllis} S.~W.~Hawking and G.~F.~R.~Ellis, {\it The Large Scale Structure of Space-Time}, (Cambridge University, Cambridge, 1973).

\bibitem{Riess} A.~G.~Riess, A.~V.~Filippenko, P.~Challis,
A.~Clocchiatti, A.~Diercks, et al., Astron.~J. {\bf 116}, 1009 (1998).

\bibitem{Perlmutter} S.~Perlmutter, G.~Aldering, G.~Goldhaber,
R.A.~Knop, P.~Nugent, et al., Astrophys.~J. {\bf 517}, 565 (1999).

\bibitem{Turner} J.~Frieman, M.~Turner and D.~Huterer, ``Dark Energy and the Accelerating Universe",
e-Print: arXiv:0803.0982 [astro-ph]. 

\bibitem{Weinberg} S.~Weinberg, Rev.~Mod.~Phys.~{\bf 61}, 1 (1989).

\bibitem{Yokoyama} A.~A.~Starobinsky and J.~Yokoyama, Phys.~Rev.~D {\bf 50}, 6357 (1994).

\bibitem{Polyakov} A.~M.~Polyakov, Nucl.~Phys.~{\bf B797}, 199 (2008).

\bibitem{Unruh} W.~G.~Unruh, Phys.~Rev.~D {\bf 14}, 870 (1976).

\bibitem{DeWitt}  B.~S.~DeWitt, in {\em General Relativity: An Einstein Centenary Survey}, S.~W.~Hawking and W.~Israel 
eds.~(Cambridge University Press, Cambridge, 1979).

\bibitem{Sasaki} M.~Sasaki, H.~Suzuki, K.~Yamamoto and J. Yokoyama, Class.~Quantum Grav.~{\bf 10}, L55 (1993).

\bibitem{GibbonsHawking}
G.~W.~Gibbons and S.~W.~Hawking, Phys.~Rev.~D~{\bf 15}, 2738 (1977).

\bibitem{BunchDavies}
T.~S.~Bunch and P.~C.~W.~Davies, Proc.~Roy.~Soc.~London~{\bf A360}, 117 (1978).

\bibitem{Tagirov} E.~A.~Tagirov, Annals~Phys.~{\bf 76}, 561 (1973).

\bibitem{Bjorken} J.~D.~Bjorken and S.~D.~Drell, {\em Relativistic Quantum Mechanics}, (McGraw-Hill, New York, 1964).

\bibitem{BirrellDavies} N.~D.~Birrell and P.~C.~W.~Davies, {\em Quantum Fields in Curved Space}, (Cambridge University, Cambridge, 1982).

\bibitem{GR} I.~S.~Gradshteyn and I.~M.~Ryzhik, {\em Tables of Integrals, Series, and Products}, A.~Jeffrey ed.\ (Academic Press, San Diego, 2000). 

\bibitem{UnruhWald} W.~G.~Unruh and R.~M.~Wald,  Phys.~Rev.~D {\bf 29} (1984). 
D29:1047-1056,1984. 

\bibitem{CHM} L.~C.~B.~Crispino, A.~Higuchi and G.~E.~A.~Matsas, Rev.~Mod.~Phys.~{\bf 80}, 787 (2008).

\bibitem{Higuchi1} A.~Higuchi, Class.~Quantum Grav.~{\bf 8}, 1983 (1991).

\bibitem{Moncrief1} V.~Moncrief, Phys.~Rev.~D~{\bf 18}, 983 (1978).

\bibitem{Moncrief2} V.~Moncrief, Gen.~Rel.~Grav.~{\bf 10}, 93 (1979).

\bibitem{Higuchi2} A.~Higuchi Class.~Quantum Grav.~{\bf 8}, 1961 (1991).

\bibitem{Losic}  B.~Losic and W.~G.~Unruh, Phys.~Rev.~D~{\bf 74}, 023511 (2006).

\bibitem{Marolfetal} D.~Marolf and I.~A.~Morrison, ``Group Averaging of massless scalar fields in $1+1$ de~Sitter", e-Print: arXiv:0808.2174 [gr-qc].

\bibitem{Marolfetal2} D.~Marolf and I.~A.~Morrison, ``Group Averaging of for de~Sitter free fields", e-Print: arXiv:0810.5163 [gr-qc]. 

\bibitem{Peskin} M.~E.~Peskin and D.~V.~Schroeder, ``An Introduction to Quantum Field Theory'', (Westview, Boulder, 1995)

\bibitem{unpublished} A.~Higuchi and Y.~L.~Lee, unpublished.
\end{thebibliography}
\end{document}